\documentclass{mn2e}
\input{psfig}
\newcommand{\etal}{{et al.~}}

\def\rmd{{\rm d}}

\def\rmf{{\rm f}}

\def\rmi{{\rm i}}

\def\rms{{\rm s}}
\def\rmt{{\rm t}}

\def\calR{{\cal R}}

\newcommand{\kmsmpc}{\>{\rm km}\,{\rm s}^{-1}\,{\rm Mpc}^{-1}}

\newcommand{\beq}{\begin{equation}}
\newcommand{\eeq}{\end{equation}}

\def\gtsima{$\; \buildrel > \over \sim \;$}
\def\ltsima{$\; \buildrel < \over \sim \;$}
\def\prosima{$\; \buildrel \propto \over \sim \;$}
\def\gsim{\lower.7ex\hbox{\gtsima}}
\def\lsim{\lower.7ex\hbox{\ltsima}}
\def\simgt{\lower.7ex\hbox{\gtsima}}
\def\simlt{\lower.7ex\hbox{\ltsima}}
\def\simpr{\lower.7ex\hbox{\prosima}}
\def\la{\lsim}
\def\ga{\gsim}
\def\lta{\la}

\newcommand{\apj}{ApJ}
\newcommand{\apjl}{ApJL}

\newcommand{\aj}{AJ}
\newcommand{\mnras}{MNRAS}
\newcommand{\aap}{A\&A}

\newcommand{\araa}{ARA\&A}

\newdimen\hssize
\newdimen\hdsize
\newcommand{\Qm}{\,Q_{\rm m}}

\newcommand{\CL}{Coulomb logarithm}

\newcommand{\dft}{$T_{\rm df}$~}
\newcommand{\AAA}{$A=1, 3.5, 10$}
\newcommand{\abest}{$A=3.5$}

\begin{document}


\title[The Dynamical Evolution of Dark Matter Subhaloes]
      {An Improved Model for the Dynamical Evolution of Dark Matter Subhaloes}

\author[Gan et al.]
       {JianLing Gan$^{1,2,3}$\thanks{Email:jlgan@shao.ac.cn},
        Xi Kang$^{4,2}$,
	Frank C. van den Bosch$^{5}$, 
        JinLiang Hou$^{1}$\\
  $^1$ Key Laboratory for Research in Galaxies and Cosmology,
    Shanghai Astronomical Observatory,\\
    Chinese Academy of Sciences, 80 Nandan RD, Shanghai,  200030, China\\
  $^2$ Max Planck Institute for Astronomy, K\"onigstuhl 17,
    69117 Heidelberg, Germany\\
  $^3$ Graduate School of the Chinese Academy of Sciences,
    No.19A, Yuquan Rd., 100049 Beijing, China \\
  $^4$ The Purple Mountain Observatory, 2 West Beijing Road, Nanjing 210008, 
    China\\
  $^5$ Department of Physics and Astronomy, University of Utah, 115
    South 1400 East, Salt Lake City, UT 84112-0830 }

\date{}

\maketitle


\begin{abstract}
  Using an analytical model, we study the evolution of subhalo,
  including its mass, angular momentum and merging time-scale. This model
  considers the dominant processes governing subhalo evolution, such as
  dynamical friction, tidal stripping and tidal heating. We find that
  in order to best match the evolution of angular momentum measured
  from N-body simulation, mass stripping by tidal force should become
  inefficient after subhalo has experienced a few passages of
  pericenter. It is also found that the often used Coulomb logarithm
  $\ln M/m$ has to be revised to best fit the merging time-scales from
  simulation.  Combining the analytical model with the Extended
  Press-Schechter (EPS) based merger trees, we study the subhalo mass
  function, and their spatial distribution in a Milky-Way (MW) type
  halo.  By tuning the tidal stripping efficiency, we can
  gain a better match to the subhalo mass function from simulation. The 
  predicted  distribution of subhaloes is found to agree with the 
  distribution of MW satellites, but is more concentrated than the 
  simulation results. The radial distribution of subhaloes depends weakly 
  on subhaloes mass at both present day and the time of accretion, but 
  strongly on the accretion time. Using the improved model, we 
  measure the second moment of the subhalo occupation distribution, and 
  it agrees well with the results of Kravtsov et al. (2004a) and 
  Zheng et al. (2005).
\end{abstract}




\section{Introduction}
\label{sec:intro}

In the popular cold dark matter model, structure (dark matter halo)
formation is processed in a hierarchical manner that small haloes form
first, and they subsequently merge to form bigger haloes.  The relics
of merging haloes are seen as the normal galaxies in clusters, or
dwarf satellites in the Milky-Way.  In the context of galaxy
formation, halo mergers play an important role, as they can
significantly affect the star formation rate and morphology of
galaxies.  It is now widely accepted that elliptical galaxies are
formed by major mergers (e.g., Toomre \& Toomre 1972), and disk
galaxies may experience preferentially minor mergers, or earlier major
mergers if any.  Thus one important aspect about galaxy formation in
the cold dark matter (CDM) scenario is to understand how and when the
mergers (halo merger) happen, how the mass and density profile of
accreted haloes evolve, and what are their final fates: merge with
central galaxies or get disrupted before sinking into the halo center.

The only  appropriate way to  study the properties of  accreted haloes
(subhaloes)   is  the   fully   dynamic-traced  simulation.    Earlier
simulations (e.g.,  Katz \& White 1993)  suffer significant resolution
effects, and  they produce the over-merging pictures  (e.g., Klypin et
al.  1999;  Moore et al.   1999).  Simulations with  higher resolution
(Springel et al.  2001; Diemand et  al.  2004; Gao et al.  2004b; Kang
et  al.  2005),  especially  the  recent ones  from  two groups  (Via
Lactea:  Diemand, Kuhlen  \&  Madau 2007a;  Aquarius:  Springel et  al.
2008) are  shown to be capable  of resolving subhalo  down to very
low  mass   and  these   simulations  converged  on   the  statistical
distributions  of subhaloes.  For example,  the subhalo  mass function
(SHMF) is found to be well described by a single power law in both low
and high-mass host haloes. Normalized  by the host halo mass, the SHMF
is universal  with  a  slight dependence  on formation time  of the
host halo (Gao et al.  2004b; Kang et al.  2005; van den Bosch et al.
2005).  Those high-resolution  N-body  simulations also  agree on  the
radial distribution of subhaloes and  it is found to be shallower than
the  dark matter  particles,  and can  be  well fitted  by an  Einasto
profile (e.g.,  Diemand et al.   2004; Springel et al.   2008).  Other
properties  of   subhaloes  are   also  discussed  but   with  diverse
conclusions, such as  the velocity bias of subhaloes,  both a positive
velocity bias (e.g., Diemand et al.  2004) and negative bias (Springel
et al.  2001) are reported.  The density profile of subhalo is rapidly
truncated, with a higher concentration, but disagreements are hold for
its  inner density  slope (Hayashi  et al.   2003; Kazantzidis  et al.
2004a; Diemand, Kuhlen \& Madau 2007b; Springel et al. 2008).

In addition to the  above statistical distributions of subhaloes, many
studies also focus on their  dynamical evolution.  It is widely agreed
that subhaloes  will sink  towards the host  halo center  by dynamical
friction and gradually lose their  mass due to tidal stripping.  There
are disagreement  on the inner  structure of subhaloes,  especially at
their  late stages  of evolution.   Hayashi et  al. (2003)  found that
subhaloes  will redistribute their  inner mass  by tidal  heating, and
become disrupted  soon once  the tidal radius  are smaller  than their
characteristic radius.   Kazantzidis et  al.  (2004a) argued  that the
inner  part of subhalo  is resistant  to tidal  shock and  subhalo can
orbit in the  host halo for a longer time.  They  pointed out that the
rapid disruption of subhaloes is unsurprised if the initial conditions
in simulations are not  constructed in equilibrium. Kazantzidis et al.
(2004b) further showed  that numerical effects can also  lead to rapid
loss  of  mass from  subhalo.   Diemand  et  al.  (2007b)  found  that
subhaloes in their  simulation can survive for longer  time even after
they have  passed the very central  part of host halo  where the tidal
force is very strong. The fate  of subhalo is even more complicated by
the presence of baryon.  A few simulations (Gnedin et al.  2004; Nagai
\& Kravtsov 2005; Macci{\`o} et  al. 2006; Weinberg et al. 2008; Dolag
et al. 2009) have shown that compared to pure dark matter simulations,
smooth particles hydrodynamic (SPH) simulations with baryon will leave
more subhaloes in the host  center as the condensation of baryon cores
makes  subhaloes more  resistant to  tidal disruption,  and  produce a
radial  distribution of  subhaloes similar  to that  observed  for the
Milky Way satellites.

Among the studies of  subhalo dynamical evolution, one important issue
is how long it takes for a subhalo to sink into the center of its host
halo. This is very important  for the model of galaxy formation as
it  determines when  the mergers  of galaxies  actually  happen.  This
time-scale  is often  called  as the  dynamical  friction time  scales
(\dft).  \dft was firstly derived by Binney \& Tremaine (1987,
hereafter BT87)  and Lacey \&  Cole (1993) based on  the Chandrasekhar
(1943)  description.  Early  simulation (Navarro  et al.   1995) found
that  the  BT87  formula  matches  well  with  the  simulation  results.
Recently Jiang et al.  (2008) and Boylan-Kolchin et al.
(2008, hereafter BK08) both find that the BT87 formula under-estimates
the merger  time-scales, and they  point out that BT87  neglected the
mass loss of subhalo during it  evolution. But even the results of 
Jiang et al.  (2008)
and BK08 differ by a factor of two, and this diversity is from various
effects.   Jiang et al.  (2008) use  cosmological  simulation with  
star formation  and
feedback, while  BK08 use controlled two-halo  merging simulation with
pure  dark   matter.   Also  they  adopt   different  definitions  for
galaxy/subhalo mergers.

Although numerical simulation  is the only proper method  to study the
dynamical evolution of subhalo, useful insight into physical processes
governing subhalo evolution can be gained from analytical model.  Base
on the  pioneer work  of Taylor \&  Babul (2001), the
analytical model was  well developed in the past  years (Benson et al.
2002; van den Bosch et al.   2005; Taylor \& Babul 2004; 2005a; 2005b;
Zentner \&  Bullock 2003; Zentner  et al.  2005).  The  model includes
the main physical processes governing subhalo evolution: gravitational
force,  dynamical   friction,  tidal  stripping,   tidal  heating  and
disruption.    Coupled   with   merger   trees   from   the   extended
Press-Schechter  (EPS)  theory, the  analytical  model  is capable  of
producing realistic  catalogue of subhaloes in given  host halo, which
can be directly compared to N-body simulation results.  Up to now most
of these  analytical works unfortunately neglect the  study of subhalo
merging time-scales. Although Taffoni et al. (2003) derived an fitting
formula for \dft, their  results are not tested against simulation
results.    BK08  recently   found  that   their  results   are  still
quantitatively  inconsistent   with  the  prediction   of  Taffoni  et
al. (2003).

In  this paper,  using an  analytical model  similar to  Zentner \etal
(2005; hereafter  Z05), we study  the dynamical evolution  of subhalo.
We investigate  the effects of  tidal stripping, Coulomb  logarithm on
the angular momentum\footnote{Throughout this paper, when referring to
  the angular  momentum, we mean it  is the angular  momentum per unit
  mass  or   the  specific  angular   momentum.}   evolution,  merging
time-scale  of subhalo.   By comparing  our model  predictions  to the
simulation results  of BK08, we proposed a  modified Coulomb logarithm
which can well reproduce the evolution of angular momentum and merging
time-scales  for   subhalo  with   different  mass  ratio   and  orbit
eccentricity.   We then combine  the analytical  model with  the Monte
Carlo merger  tree to  produce subhalo catalogue  in a Milky  Way (MW)
type halo, and compare the model predictions to both N-body simulation
results and observed distribution of  satellites in the Milky Way.  In
Section~\ref{sec:model}  we  present   the  main  ingredients  of  the
analytical    model    and    show    the   model    predictions    in
Section~\ref{sec:results}. In  Section~\ref{sec:HOD} we further examine
our  model with  the  halo occupation  distribution  of subhaloes.  In
section~\ref{sec:discuss} we discuss the radial distribution of subhaloes,
and we briefly conclude our model in Section~\ref{sec:conc}.

Throughout this paper we adopt a flat $\Lambda$CDM cosmology with the
following cosmological parameters: $\Omega_{\rm m} =0.25$,
$\Omega_{\Lambda}=0.75$, $h=H_0/(100 \kmsmpc) = 0.73$, $\Omega_{\rm
  b}=0.04$, $n_{\rm s} = 0.951$ and $\sigma_8=0.9$. 

\section{The Model}
\label{sec:model}

In  this  section  we present  the  model  for  the evolution  of  the
population  of dark  matter subhaloes.  The first  part  describes the
merger history  (i.e. mass  assembly) of the  host halo, which  can be
obtained  using either the  EPS theory  or $N$-body  simulations.  The
second  part  describes  the  dynamical evolution  of  subhaloes,  and
includes  orbit  integration in  the  presence  of dynamical  friction
combined with  tidal stripping  and heating. Our  model is  similar to
that of  Z05, but  there are  also a few  differences.  We  employ the
well-calibrated code of Parkinson et  al.  (2008) to construct the EPS
merger trees. This code is a significant improvement  over the Somerville
\& Kolatt (1999) implementation used by Z05. In addition, we calibrate
our model using detailed numerical simulations on the evolution of the
orbital angular momentum of  subhaloes. As we will demonstrate,
this kind of calibration is  far more constraining than using the mass
function or velocity function of subhaloes. Finally, we (i) use a more
detailed, empirical  treatment of tidal heating, based  on the results
of high resolution numerical  simulations, (ii) investigate the impact
of changes in  the Coulomb logarithm used in  the analytical treatment
of dynamical  friction, and (iii)  consider a different  treatment for
the tidal mass loss of subhaloes.

In what follows, we use $m$ and $M$ to denote the instantaneous masses
of subhalo and host halo, respectively. Unless stated otherwise,
we consider it understood that both $m$ and $M$ are functions of time.  
Then we use the symbol $\mu$ to refer to the mass ratio  between subhalo
and host halo, i.e., $\mu=\mu(t)=m(t)/M(t)$, without writing the
time-dependence explicitly. We use $\mu_{\rmi}$, $\mu_{\rmf}$ to refer to the
initial mass ratio at the time of accretion ($t_{\rm acc}$ or $z_{\rm acc}$) 
and the final mass ratio at present day ($z=0$), i.e.,
$\mu_{\rmi}=m(t_{\rm acc})/M(t_{\rm acc})$, $\mu_{\rmf}=m(z=0)/M(z=0)$, respectively.

\subsection{Merger Trees}
\label{subsec:tree}
\begin{figure}
\centerline{\psfig{figure=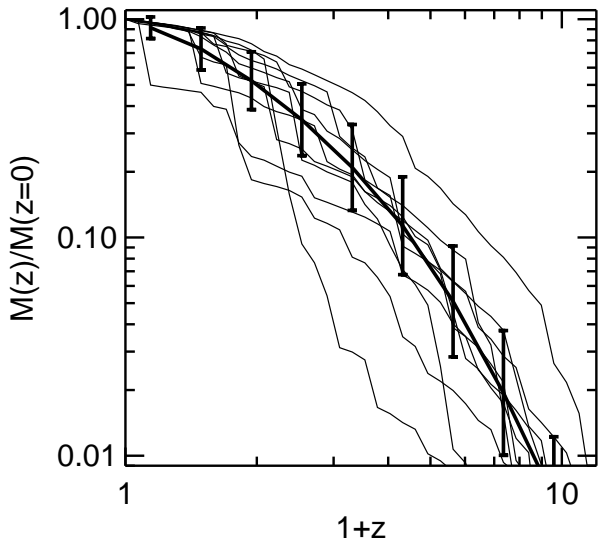,width=0.95\hsize}}
\caption{The $10$ random mass accretion histories (MAHs, thin lines) for 
      	 the mass of the main haloes as function of time, for a MW
	 sized halo with $M(z=0) = 1.77\times 10^{12} h^{-1}M_{\odot}$.
	 The thick line with error bars show the average MAH and 
	 $1-\sigma$ scatters from $100$ such realizations. }
\label{fig:MAH}
\end{figure}

The backbone for modeling the population of dark matter subhaloes is
the merger history of the host halo, which describes when each subhalo
is accreted, and what its mass is at accretion.  Halo merger histories
can  be  obtained  using    $N$-body  simulations,  or  in  a
semi-analytical  fashion from  the  EPS formalism  (Bond \etal  1991;
Lacey  \& Cole 1993).  Here we  adopt the  latter approach by employing the
open-source code of  Parkinson et al. (2008) to  generate the assembly
histories of MW  sized haloes with a $z=0$  mass of $M(z=0)=1.77\times
10^{12} M_{\odot}$ that is close to the $z=0$ mass of the host halo in
``Via  Lactea''  and  ``Aquarius''  simulations.  We  construct  $100$
independent merger-tree realizations, each  with a resolution of $10^8
M_{\odot}$.  In Fig.~\ref{fig:MAH} we show an example of the halo mass
accretion  history (MAH), which  is defined  as the  mass of  the most
massive progenitors at  each redshift from the merger  tree.  The thin
lines are $10$  random MAHs of the MW type halo  in our adopted cosmology. The
thick  line with  error bars  is the  average MAH  and  its $1-\sigma$
scatters (standard deviation) from $100$ such realizations.

The main branch of the merger tree is defined as the trajectory of the
most massive progenitors starting from the $z=0$ halo.  In our study
we consider only those haloes that are directly accreted onto the main
branch, not accounting for any of their subhaloes (which would give
rise to sub-subhaloes).  As shown by Yang, Mo \& van den Bosch (2009),
sub-subhaloes (and higher-order substructures) only contribute a small
fraction to the total substructure mass function (see also Giocoli et
al. 2010).

For each halo in the merger trees, its virial radius, $r_{\rm vir}$, is
defined as the radius within which the mean mass density is 
$\Delta_{\rm c}(z)$ times the critical density of the universe at 
redshift $z$, where
\begin{equation}
\label{eq:delta}
\Delta_{\rm c}(z) = 18 \pi^2 + 82 x - 39 x^2\,
\end{equation}
with $x = \Omega_{\rm m}(z) - 1$ (Bryan \& Norman 1998). We assume
that the host haloes are spheres with a density distribution
given by the NFW profile (Navarro \etal 1997).  The corresponding
concentration parameter, $c$, is set using the median relation between
$c$ and halo mass $M$ of Neto \etal (2007), which is given by
\begin{equation}
\label{eq:c_m}
c(M,z) = {4.67\over 1+z} \, 
\left[\frac{M(z)}{10^{14}h^{-1}M_{\odot}} \right]^{-0.11} \, ,
\end{equation}
where the dependence on redshift is taken from Bullock \etal (2001).
Dark matter subhaloes are assumed to have a similar NFW profile at
their time of accretion (i.e., at the time they become a subhalo), but
as described below (see Section~\ref{subsubsec:tides}), this density 
profile is subsequently modified due to tidal heating.

\subsection{Dynamical Evolution of Subhaloes}
\label{subsec:dynamics}

\subsubsection{Orbital Parameters}
\label{subsubsec:orbparam}

The first step for the dynamical evolution of dark matter
subhaloes is to assign their orbital parameters at the time of
accretion, $t_{\rm acc}$. We follow Z05 and draw the initial orbital
energy and angular momentum from distributions that have been obtained
from $N$-body simulations. In particular, we assume that each subhalo
starts its orbit at the virial radius, $r_{\rm vir}$, of the host halo
at the time of accretion with an orbital energy equal to that of a
circular orbit of radius $\eta r_{\rm vir}$, where $\eta$ is drawn
randomly from a uniform distribution between $[0.6,1.0]$ (see
Z05). The initial specific angular momentum is parameterized as
$j_{\rm init} = \varepsilon j_{\rm c}$, where $j_{\rm c}$ is the
specific angular momentum of the circular orbit mentioned above and
$\varepsilon$ is called the orbital circularity (note that $0 \leq
\varepsilon \leq 1$). Several studies (e.g., Benson 2005; Tormen 1997;
Z05; Khochfar \& Burkert 2006; Jiang et al. 2008) have measured the 
distribution of
$\varepsilon$, all reporting similar results.  Here we use the
distribution obtained by Jiang et al.  (2008):
\begin{equation}
\label{eq:f_epson}
f(\varepsilon) = 2.77 \varepsilon^{1.19} 
\left(1.55-\varepsilon\right)^{2.99} \;.
\end{equation}

\subsubsection{Dynamical Friction}
\label{subsubsec:dynfric}

We treat subhalo as test particle in the orbital evolution.
In addition to the (radial) force due to the gravitational
potential of the spherical NFW host halo, subhalo experience an
effective force due to `dynamical friction' caused by the
gravitational interaction between subhalo and the background
`field' particles that make up the host halo. Chandrasekhar (1943)
showed that if the distribution of background particles is infinite and
homogeneous, one can obtain an analytical expression for the dynamical
friction force by considering the cumulative effect of many {\it
  uncorrelated} two-body interactions\footnote{By considering the
  interactions to be uncorrelated, one effectively ignores the
  self-gravity of the field particles.} between the subject mass (in
our case the subhalo) and the individual field particles. This is
known as the Chandrasekhar dynamical friction force, which is given by
\begin{equation}
\label{DForig}
{\bf F}_{\rm df} = -4\pi \left({Gm \over v_{\rm orb}}\right)^2 
 \ln\Lambda \, \rho(<v_{\rm orb}) {{\bf v}_{\rm orb}\over v_{\rm orb}}\,.
\end{equation}
Here $m$ and $v_{\rm orb}$ are the mass\footnote{Note that the mass 
entering to the dynamical friction may not be the same as the 
bound mass of subhalo (e.g., Fellhauer \& Lin 2007). Here, an 
`effective' mass contributing to the dynamical friction is modeled. } 
and orbital velocity of the
subhalo, $\ln\Lambda$ is the \CL{}, and $\rho(<v_{\rm orb})$ is the
density of the particles in the host halo that have a speed less than
the velocity of the subhalo. The \CL{} is introduced to avoid divergence that
arises from the assumption of an infinite, homogeneous sea of field
particles.

Similar to the frictional  drag in fluid mechanics, dynamical friction
exerts a  force always  opposite to the  motion. However,  contrary to
hydrodynamic  friction, which  always increases  in strength  when the
velocity  increases, the  drag due  to dynamical  friction has  a more
complicated dependence  on velocity. While $F_{\rm  df} \propto v_{\rm
  orb}$  in  the  low  $v_{\rm orb}$-limit,  similar  to  hydrodynamic
friction, one  has that $F_{\rm  df} \propto v_{\rm orb}^{-2}$  in the
high $v_{\rm orb}$-limit.  In what follows we assume  that the `field'
particles  that make  up the  host  halo follow  a locally  Maxwellian
velocity distribution.  In that case,  Equation~(\ref{DForig}) reduces
to
\begin{equation}
\label{eq:f_df}
{\bf F}_{\rm df} = -4\pi \left({Gm \over v_{\rm orb}}\right)^2
 \ln\Lambda \, \rho(r) \left[{\rm erf}(X)-{2X \over \sqrt{\pi}}
 e^{-X^2}\right] {{\bf v}_{\rm orb}\over v_{\rm orb}}\,
\end{equation} 
(BT87), where $X=v_{\rm orb}/[\sqrt{2}\sigma(r)]$
with $\sigma(r)$ the local, one-dimensional velocity dispersion of the
host halo at radius $r$, which can be solved using the Jeans equation
(BT87; Cole \& Lacey 1996)
under the assumption that the stress tensor is isotropic\footnote{see
Zentner \& Bullock (2003) for a useful fitting function.}. 

\subsubsection{Coulomb Logarithm}
\label{subsubsec:coullog}

As shown by White (1976), in the case of an extended subject mass,
as is the case for our subhalo, one has that
\begin{equation}
\label{eq:CL_white}
\ln\Lambda = {1 \over m^2} \int_0^{b_{\rm max}} I^2(b) \, b^3 \, 
\rmd b
\end{equation}
where
\begin{equation}
\label{eq:Ib_white}
I(b) = \int_b^{\infty} {m(r) \, \rmd r \over r^2 \, (r^2 - b^2)^{1/2}}
\end{equation}
with $m(r)$ the subhalo mass profile. Here $b_{\rm max}$ is the
maximum impact parameter considered, which is introduced in order to
avoid divergence. This divergence, however, arises because
Equation~(\ref{DForig}) is based on the (unrealistic) assumption of a
homogeneous and infinite medium. In our case of a subhalo orbiting in
a host halo, a logical value for $b_{\rm max}$ may appear to be the
size of the host system, which is indeed what is often
adopted. However, it is important to realize that strictly speaking
Equation~(\ref{DForig}) is not valid for this case, and that there is no
`correct' value for the \CL{}. Hence, different forms for $\ln\Lambda$
have been adopted in the literature.  Some authors treat the \CL{} as
a constant (Velazquez \& White 1999; Taylor \& Babul 2001; Jardel \&
Sellwood 2009). Others claims that this yields a dynamical friction
time, $T_{\rm df}$, defined as the timescale on which the subject mass
looses its orbital angular momentum, that is too short, and advocate
instead that $\ln\Lambda$ has to be time-dependent (e.g., Colpi \etal
1999; Hashimoto \etal 2003). A widely used form for the \CL{} is
$\ln(M/m)$ or $\ln(1+M/m)$, where $M$ and $m$ are the instantaneous
(time-dependent) mass of the host and subhalo, respectively
(e.g. BK08; Jiang et al. 2008).  In this paper we will consider two 
different forms
for the \CL{}: $\ln\Lambda=C$ and $\ln\Lambda=-\ln\mu+C$. As we
will show, both forms yield equally satisfactory results (when
compared to numerical simulations), as long as $C$ is allowed to vary
with the initial orbit parameters, $\eta$ and $\varepsilon$, and the
initial (i.e., at the time of infall) mass ratio $\mu_{\rmi}$.

\subsubsection{Orbit Integration}

We integrate the orbits [${\bf x}(r, \theta)$] of subhalo by treating it as test
particle. The equation of motion for a subhalo of mass $m$ is given
by:
\begin{equation}
\label{eom}
{\rmd^2{\bf x} \over \rmd t^2} = -{G M(<r) \over r^2} \, 
{{\bf r}\over r} + {{\bf F}_{\rm df} \over m}\,
\end{equation}
with $M(<r)$ the mass of the host halo inside of radius $r$, and 
${\bf  F}_{\rm df}$ the dynamical friction force given by
Equation~(\ref{eq:f_df}).  The equation of motion is solved using a
fifth-order Cash-Karp Runga-Kutta method. During time-steps in which
the mass of the host halo increases (due to the accretion of a new
subhalo), we recompute the mass distribution and potential of the host
halo, always under the assumption that the host halo has a NFW shape
with the $c(M)$ relation as given by Equation~(\ref{eq:c_m}). We assume
that the orbital angular momentum of a subhalo is conserved when the
mass of the host halo increases; the only mechanism by which the subhalo
is assumed to lose orbital angular momentum is dynamical friction.

\subsubsection{Tidal Stripping and Heating} 
\label{subsubsec:tides}

When a subhalo orbits its host halo, it loses mass due to tidal
stripping. The tidal radius, $r_{\rm t}$, is the radius in subhalo
where the external differential (tidal) force from the host halo 
exceeds the binding force of the subhalo, and is approximated by
\begin{equation}
\label{eq:rt}
r_{\rm t}^3 = \frac{G m(<r_{\rm t})}
{\omega^2 + G\left[ 2M(<r)/r^3 - 4\pi\rho(r) \right] } \,,
\end{equation}
 with $\omega$  the angular  speed of the  subhalo and  $\rho(r)$ the
density profile of the host  halo (von Hoerner 1957; King 1962; 
Taylor \&  Babul 2001).
The subhalo mass outside $r_{\rm t}$ becomes unbound and is ultimately
stripped.    It    should    be    pointed    out,    however,    that
Equation~(\ref{eq:rt}) is only a crude approximation. First of all, in
the case of  non-circular orbits the concept of a  tidal radius is not
well  defined.  Secondly,  even  in  the case  of  point  masses,  the
two-dimensional surface  along which $\rmd^2  r_{\rm t}/\rmd t^2  = 0$
(i.e., zero-velocity surface; BT87) is not spherical, and so cannot be
characterized by a  single radius. And finally, Equation~(\ref{eq:rt})
ignores the orbital motion of particles within the subject mass. This,
among  other   effects,  gives  rise  to  scatter   in  $\omega$,  and
effectively  introduces  some non-zero  `thickness'  to  the shell  of
particles for which the internal and tidal forces balance.

Because of these uncertainties, numerical simulations show that many
particles remain bound even though they lie beyond the tidal limit of
Equation~(\ref{eq:rt}) when the subhalo is near pericenter (e.g., Diemand
\etal 2007b). Fellhauer \& Lin (2007) also showed that the previously 
stripped material can contribute to the dynamical friction and affect mass 
loss from the subhalo. This has resulted in uncertainties regarding 
how best to model tidal stripping.  In particular, different studies adopt
different time scales for tidal stripping, $T_{\rm strip}$, defined by
\begin{equation}
\label{eq:strip}
{{\rm d} m \over {\rm d} t} = -\frac{m(>r_{\rm t})}{T_{\rm strip}} \,.
\end{equation}
Whereas Taylor \&  Babul (2001) simply assumed that $T_{\rm strip}$ is 
equal to the
instantaneous orbital time $T_{\rm orb} \equiv 2\pi/\omega$, Z05 and
Diemand \etal (2007b) inferred stripping time-scales that are $3.5$
and $6$ times shorter, respectively. In order to parameterize this
uncertainty we adopt
\begin{equation}
\label{eq:Adef}
T_{\rm strip} \equiv {T_{\rm orb} \over A} \,,
\end{equation}
which we use in combination with Equations.~(\ref{eq:rt})
and~(\ref{eq:strip}) to describe mass loss due to tidal
stripping. Here $A$ is the tidal stripping efficiency parameter, which
we tune using detailed numerical simulations (see Section~\ref{subsec:subpop}
below). Note that Taylor \&  Babul (2001), Z05 and Diemand \etal (2007b) 
used or
advocated $A \simeq 1.0$, $3.5$ and $6.0$, respectively.

Using numerical $N$-body simulations, it has been shown that tidal
heating causes subhaloes to expand and to reduce their inner mass
profile (e.g., Hayashi \etal 2003; Kravtsov \etal 2004b). Hayashi
\etal (2003) introduced a modified NFW profile to describe the density
distribution of a tidally heated subhalo according to
\begin{equation}
\label{eq:heating}
\rho(r) = \frac{f_\rmt}{1+(r/r_{\rm te})^3} \rho_{\rm NFW}(r)\,.
\end{equation}
Here $\rho_{\rm NFW}(r)$ is the original NFW density profile of the
subhalo at the time of infall, $r_{\rm te}$ is the `effective' tidal
radius that describes the outer cutoff imposed by the tides, and
$f_\rmt$ describes the reduction in the central density of the
subhalo. As shown by Hayashi \etal (2003), these are well-fit by
\begin{equation} 
\lg \frac{r_{\rm te}}{r_\rms} = 1.02 + 1.38\Qm + 0.37\Qm^2\,,
\end{equation}
and
\begin{equation}
\lg f_\rmt=-0.007+0.35 \Qm + 0.39 \Qm^2 + 0.23 \Qm^3\,.
\end{equation}
Here $\Qm = \lg[m(t)/m(t_{\rm acc})]$ is the logarithm of the
remaining fraction of subhalo mass, and $r_\rms$ is the scale radius
of the NFW profile at the time of accretion. Both $f_\rmt$ and $r_{\rm
  te}$ decrease with time while a subhalo is losing mass.

We caution that this `empirical' treatment of tidal heating is subject
to some debate. In particular, Kazantzidis \etal (2004a) have argued
that the simulation used by Hayashi \etal (2003) was not set-up in
equilibrium, and that this has resulted in a tidal mass loss rate that
is too high. Unfortunately, lacking a more reliable description of how
tidal heating impacts the density profiles of subhaloes, we use
Equation~(\ref{eq:heating}) despite these potential problems.

\subsubsection{Disruption and Cannibalism}
\label{subsubsec:fate}

As a subhalo is being exposed to tidal stripping and heating, it may
reach a point at which it is disrupted, i.e., at which no significant
amount of matter remains gravitationally bound into a single
object. Alternately, depending on the dynamical friction time, a
subhalo may either sink all the way to the center of the host halo's
potential well (i.e., lose all its orbital angular momentum) or
continue to orbit as a subhalo, if the mass ratio $m(t)/M(t)$ is such
that dynamical friction is negligible.

Whether and when subhaloes are tidally disrupted is still being
debated. Testing this with numerical simulations is complicated by the
fact that simulations are always subject to numerical artifacts due to
limited mass and force resolution. Using $N$-body simulations, Hayashi
\etal (2003) found that subhaloes are disrupted once their tidal
radius $r_\rmt$ becomes smaller than $\sim 2 r_\rms$. Motivated by
these findings, Taylor \& Babul (2004) and Z05 included tidal
disruption in their semi-analytical models.  They considered a subhalo
to become tidally disrupted once its mass becomes less than its
initial mass (i.e. at accretion) within a radius $f_{\rm dis} r_\rms$,
with $f_{\rm dis}=0.1$ (Taylor \& Babul 2004) and $f_{\rm dis}=1.0$
(Z05), respectively.  Using the luminosity function of Milky Way
satellites, Macci{\`o} \etal (2009) argued that $0.1 \lta f_{\rm dis}
\lta 0.5$, while Wetzel \& White (2010), using a wide variety of
observational constraints on satellite galaxies, conclude that
subhaloes are disrupted once their bound mass drops below $\sim 2$
percent of its mass at infall, corresponding to $f_{\rm dis} \sim 0.3$
for a NFW profile with a concentration $c=10$.  Hence, typical values
for $f_{\rm dis}$ that have been adopted in the literature cover the
entire range $0.1 \lta f_{\rm dis} \lta 2$; as nicely shown by Wetzel
\& White (2010), varying $f_{\rm dis}$ by this amount has a huge
impact on the radial number density distribution of surviving
subhaloes, with smaller $f_{\rm dis}$ resulting in a more concentrated
profile.

There have also been a number of studies that have argued that subhalo
disruption is actually extremely rare. In particular, a number of
numerical simulations with very high spatial resolution have shown
that subhaloes are remarkably resilient to disruption by tidal shocks
(e.g., Kazantzidis \etal 2004a; Bullock \& Johnston 2005;
Pe{\~n}arrubia \etal 2008; Springel \etal 2008). Kanzantzidis \etal
(2004a) argued that the initial conditions of the simulation of
Hayashi \etal (2003) were not in equilibrium, which is likely to have
caused a subhalo disruption rate that is too high. Using the ``Via
Lactea'' simulation, Diemand \etal (2007b) found that only very few
subhaloes get disrupted; In their simulation 97\% of the subhaloes
identified at $z=1$ were still present at $z=0$, and even subhaloes
with $r_{\rmt} < 0.2 r_\rms$ were found to survive.  Finally, BK08
also found that subhaloes survive as bound entities up to the point of
having lost all their orbital angular momentum (private
communication).

Motivated by these high-resolution simulations, we assume that
subhaloes are never disrupted. Rather, when a subhalo has lost all its
orbital angular momentum, we consider it `cannibalized' by (or
`merged' with) the host halo, and we remove the subhalo from our
sample\footnote{Note that some authors consider subhaloes `merged' or
  `cannibalized' when its separation from the host center is smaller
  than some fiducial radius (e.g., Kravtsov et al. 2004a;
  Z05). However, we consider our definition, based on the complete
  loss of orbital angular momentum, more realistic (see also
  BK08).}. The merging time scale, $T_{\rm df}$, is defined as the
time interval between accretion and merging of a subhalo. We consider
two cases for subhalo mass loss due to tidal stripping.  In the first
case (hereafter Model M1), subhaloes can lose mass continuously until
they are cannibalized by their host halo. In the second case
(hereafter Model M2), tidal stripping is `turned off' after a subhalo
has experienced two pericentric passages.  This is motivated by
numerical simulations, which suggest that subhaloes only experience
significant mass loss during their first two orbital periods (e.g.,
Diemand \etal 2007b). As we will show below (see
Section~\ref{subsec:tuning}), these two different treatments of
tidal stripping yield very different predictions regarding the
evolution of the orbital angular momentum of subhaloes.

\section{Comparison with Numerical Simulations}
\label{sec:results}
\begin{figure*}
\centerline{\psfig{figure=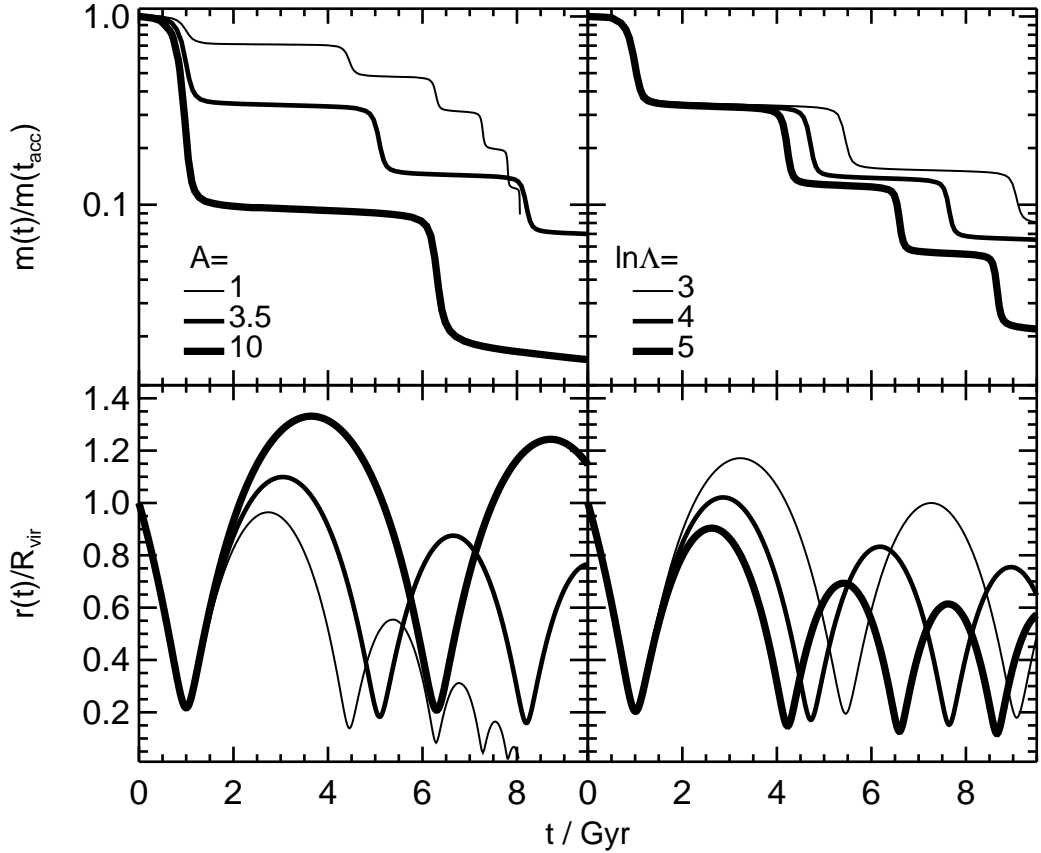,width=0.8\hsize}}
\caption{The evolution of mass and halo-centric radius for a subhalo
  with $\mu_{\rmi} = 0.05$, $\varepsilon = 0.5$, and $\eta = 1.0$ in model
  M1.  In the left-hand panels, $\ln \Lambda =- \ln \mu$ and we vary 
  the efficiency  $A$ of tidal stripping, as indicated. In the 
  right-hand panels,  $A=3.5$ and we vary the value of the Coulomb 
  logarithm, as  indicated.}
\label{fig:modelA}
\end{figure*}

We now turn to a detailed comparison of our analytical model with
numerical simulations. After exploring how the orbital evolution of a
subhalo depends on the tidal stripping efficiency, $A$, and the
Coulomb logarithm, $\ln\Lambda$, we tune these parameters by fitting
the merging time-scales and the evolution of orbital angular momentum
of subhaloes to the controlled, high-resolution numerical simulations
of BK08. These simulations follow the orbital evolution of individual
subhaloes of different mass and with different orbital properties in a
host halo of fixed mass $M(t) = M$, and are ideally suited to tune our
model parameters.

Subsequently, we use our model to compute the mass function and radial
number density distribution of subhaloes in a MW type host halo, which
we compare to numerical simulations and observational constraints from
satellite galaxies in the MW.

\subsection{Tuning the Tidal Stripping Efficiency and Coulomb Logarithm}
\label{subsec:tuning}

Fig.~\ref{fig:modelA} shows the evolution of the mass (upper panels)
and halo-centric radius (lower panels; in units of the virial radius
of the host halo) of a subhalo with $\mu_{\rmi} = 0.05$, $\varepsilon =
0.5$ and $\eta=1.0$ in the model M1.  In the left-hand panels the Coulomb
logarithm is $\ln \Lambda =- \ln \mu$ and we vary the efficiency of tidal stripping,
expressed by the parameter $A$ (see Equation~[\ref{eq:Adef}]), as
indicated.  Clearly, for larger values of $A$ (i.e., more rapid mass
loss), the effect of dynamical friction is reduced, and the orbital
decay slows down drastically. With decreasing dynamical friction, subhalo can 
also travels to higher halo-centric radius (lower panel).
\begin{figure}
\centerline{\psfig{figure=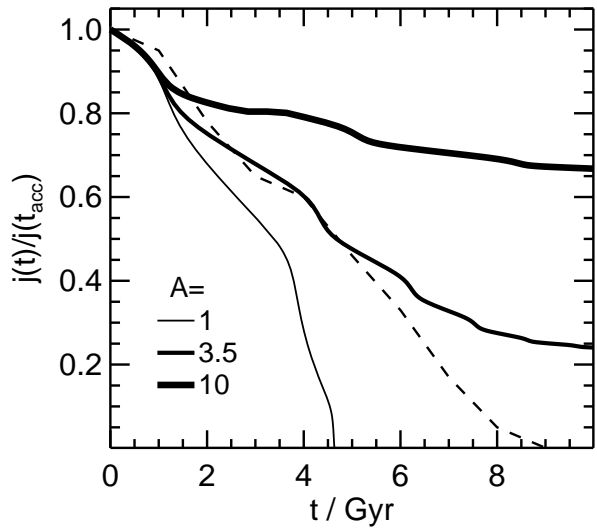,width=0.95\hsize}}
\caption{The evolution of angular momentum with $\mu_{\rmi} = 0.1$,
  $\varepsilon = 0.65$, $\eta = 1.0$, $\ln \Lambda =- \ln \mu$ from 
  model M1.   The solid lines are the model results with different 
  tidal stripping efficiency (\AAA). The dashed 
  line shows   the simulation   result  from BK08.}
\label{fig:badj1}
\end{figure}

In the right-hand panels, we keep $A$ fixed as $3.5$ and vary the Coulomb logarithm,
as indicated. Typically, a lower value for $\ln\Lambda$ results in
dynamical friction being less efficient (cf. Equation~[\ref{eq:f_df}]).
This in turn implies a reduced mass loss, because the subhalo
experiences fewer pericentric passages and, on average, more orbits at
larger halo-centric radii where the tidal forces due to the host halo
are weaker.

We now  turn to a detailed  comparison with the  simulation results of
BK08. To that extent, we use  the same initial conditions, such as the
density  profiles of subhalo  and host  halo, orbital  circularity and
orbital  energy.  The  dashed line  in Fig.~\ref{fig:badj1}  shows the
evolution  of   $j(t)$  for  a   subhalo  with  $\mu_{\rmi}   =  0.1$,
$\varepsilon = 0.65$, and $\eta = 1.0$ in the simulation of BK08.  The
solid lines  show the predictions  from our model M1,
for three different  values of $A$, as indicated.   In all three cases
we  have used $\ln  \Lambda =-  \ln \mu$.   Clearly, the  evolution of
orbital angular  momentum is  a strong function  of the  efficiency of
tidal stripping,  $A$.  Larger values of  $A$ result in  lower rate of
angular  momentum loss  (i.e., a  longer merging  time  scale, $T_{\rm
  df}$),  simply because  a  less massive  subhalo experiences  weaker
dynamical friction.  Whereas our  model with $A=3.5$ matches $j(t)$ in
the BK08  simulation reasonably well for  the first $\sim  5$ Gyr, the
predicted evolution in orbital angular momentum at later stages is too
weak.  We have  experimented with  different values  of $A$  but where
unable to obtain a satisfactory  match to the $j(t)$ in the simulation
of BK08.
\begin{figure}
\centerline{\psfig{figure=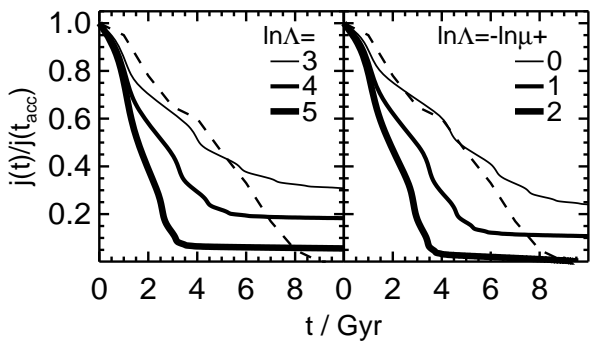,width=0.95\hsize}}
\caption{  As Fig.~\ref{fig:badj1}, but  with $A$  is fixed  as 3.5
   (constrained  from  subhalo mass  function,  shown  in 
   Fig.~\ref{fig:subpop}).  The figure shows the effects of varying
   Coulomb logarithm. Note the predicted slow evolution at later 
   stages.}
\label{fig:badj2}
\end{figure}

In Fig.~\ref{fig:badj2}, we investigate the impact of changing the
Coulomb logarithm.  In the left-hand panels we keep $\ln\Lambda$ fixed
at some constant values (as indicated), while in the right-hand panels
we adopt $\ln\Lambda = -\ln \mu + C$ for three different values of
$C$ (as indicated). In all case $A$ is fixed as $3.5$, which, as we will see in
Section~\ref{subsec:subpop}, yields a subhalo mass function that is in best
agreement with numerical simulations. Although different Coulomb
logarithms have a significant impact on the evolution of the orbital
angular momentum, we were unable to find a form for $\ln\Lambda$ for
which we could satisfactorily match the simulation results of BK08, even
if we kept $A$ a free parameter as well.  The problem is that the
model typically predicts a decline in the angular momentum loss rate,
$\calR \equiv -{\rm d}j/{\rm d}t$, while the simulation results have
$\calR \sim {\rm constant}$ during the entire evolution, up to the
point of being cannibalized by the host (i.e., when $j = 0$).  The
only exception is when the tidal stripping efficiency $A$ is very low,
in which case the merging time scale, $T_{\rm df}$, is much too short.
The culprit for this discrepancy is the continued mass loss due to
tidal stripping. This has motivated us to consider a modified model
(model M2) in which tidal stripping is inefficient after two
pericentric passages. As already mentioned above, this is not an
entirely ad-hoc modification, as it has support from the ultra-high
resolution ``Via Lactea'' simulation (Diemand \etal 2007b).
\begin{figure}
\centerline{\psfig{figure=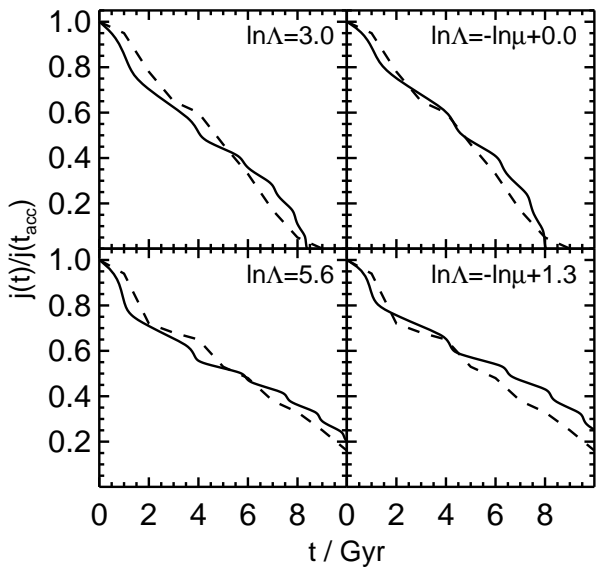,width=0.95\hsize}} 
\caption{The evolution of angular momentum in model M2, in which tidal
  stripping is stopped after subhalo has gone through two pericenter
  passages.  The upper panels are evolutions with $\mu_{\rmi} = 0.1$,
  $\varepsilon = 0.65$ and $\eta = 1.0$. Lower panels are for subhalo
  with $\mu_{\rmi} = 0.05$, $\varepsilon = 0.46$, $\eta = 1.0$.  The two
  forms of \CL{} that $\ln\Lambda=C$ and $\ln\Lambda=-\ln\mu+C$
  are used in left and right panels, respectively. The solid and
  dashed lines indicate the model and simulation (BK08) predictions.
  Better agreement between them can be obtained with appropriate
  $\ln\Lambda$.}
\label{fig:goodj1}
\end{figure}

Fig.~\ref{fig:goodj1}  shows  the  evolution  of the  orbital  angular
momentum of  two different subhaloes from  the model M2;  in the upper
panels $\mu_{\rmi}  = 0.1$,  $\varepsilon = 0.65$,  and $\eta  = 1.0$,
while the  lower panels correspond to  a subhalo with  the same $\eta$
but  with $\mu_{\rmi}  = 0.05$  and $\varepsilon  = 0.46$.  The dashed
lines indicate  the results  from the simulations  of BK08,  while the
solid lines  correspond to our model M2  with $A = 3.5$,  and with the
Coulomb logarithm tuned to best  match the BK08 results.  In the left-
and right-hand panels we considered $\ln\Lambda = C$ and $\ln\Lambda =
-\ln \mu  + C$, respectively,  where $C$ is  a constant.  In  all four
cases, the fit to the $j(t)$ of BK08 is fairly satisfactory.  Clearly,
model M2 gives much better fit to the simulation results, and we adopt
this model throughout the remainder of this paper (unless specifically
stated otherwise).

Note, though, that the best-fit value of $C$ (indicated in each panel)
is different in each case. Both the model with $\ln\Lambda = C$ and
that with $\ln\Lambda = -\ln\mu + C$ yield equally satisfactory
results. In what follows we will only consider the latter, since we
believe it to be the more physical one. What remains to be done,
however, is to characterize how $C$ depends on the mass and orbital
properties of the subhalo. Using a suite of numerical simulations,
BK08 derived a fitting formula for the merging time scale, $T_{\rm
  df}$, as a function of the mass, $m$, the orbital circularity
$\varepsilon$, and the orbital energy, $\eta$, of the subhalo at
accretion. We use this fitting formula to constrain $C =
C[\mu_{\rmi},\eta,\varepsilon]$. After some experimenting, we finally
adopted the following form for the Coulomb logarithm:
\begin{equation}
\label{eq:fitting}
\ln\Lambda = -\ln\mu + c_1 \mu_{\rmi}^{c_2} \eta^{c_3}
\exp[c_4 \varepsilon]  + c_5\,.
\end{equation}
Fitting the merging time scales in our semi-analytical model to those
listed in Table~1 of BK08, we obtain: $c_1 = 1.04$, $c_2 = -0.64$, 
$c_3 = 0.72$, $c_4 = -3.02$, and $c_5=-0.75$. 

\begin{figure*}
\centerline{\psfig{figure=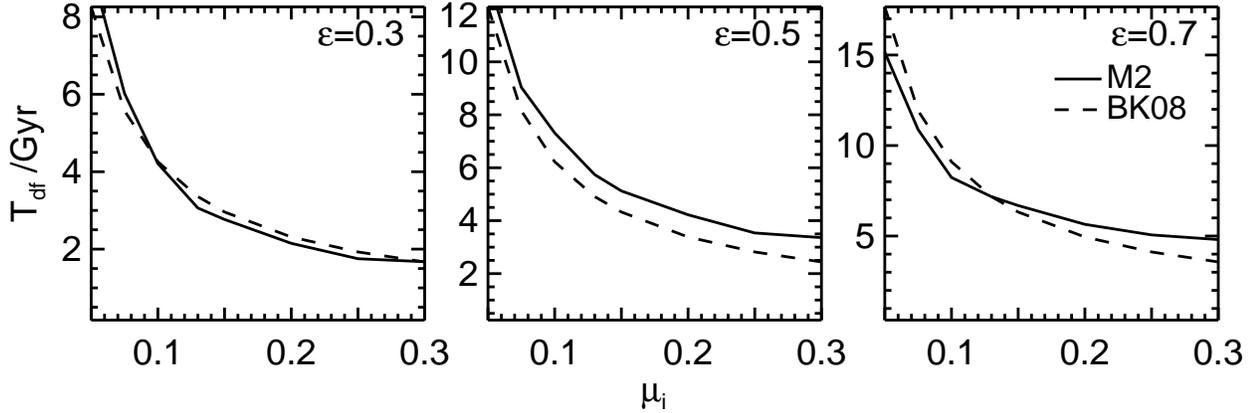,width=0.95\hsize}}
\caption{Subhalo merging time-scales. As  in BK08, subhalo merges with
  central halo  once it  loses all of  its angular momentum  $j$.  The
  three panels are for different orbital circularity $\varepsilon=0.3,
  0.5, 0.7$.  The solid lines  are from model M2 with $\eta = 1.0$, 
  \abest{} and $\ln\Lambda=-\ln\mu+C$ given by Equation~(\ref{eq:fitting}).
  The  dashed  line  shows  the  fitting
  formula from simulations of BK08. }
 \label{fig:tmerge}
\end{figure*}

\begin{figure}
\centerline{\psfig{figure=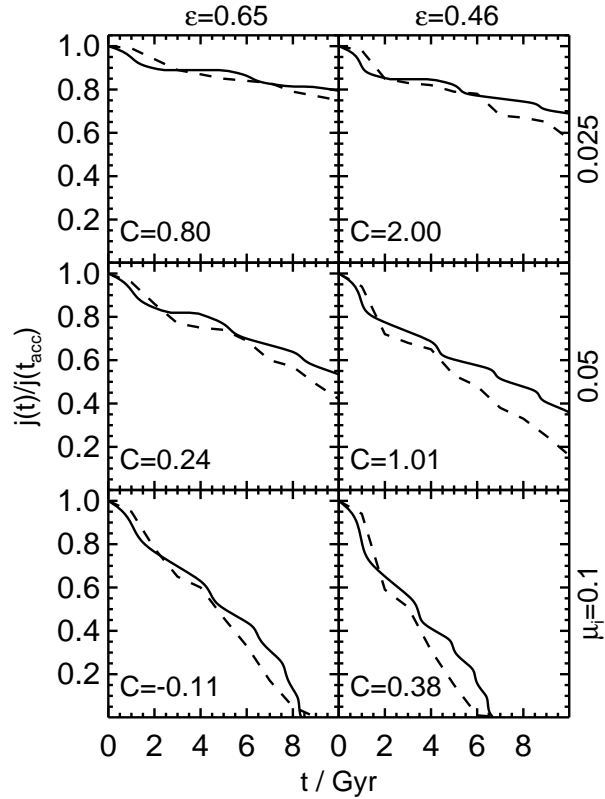,width=0.95\hsize}}
\caption{The evolution  of angular momentum for  subhalo.  The dashed
   lines  are from  simulation  by  BK08, and  solid  lines are  model
   predictions. Here the Coulomb logarithm is from 
   Equation~(\ref{eq:fitting}), and $C$  is  labeled in  each panel.
   It  shows that  our revised Coulomb logarithm can  well describe the
   evolution of angular  momentum for subhaloes with different mass 
   ratio, orbit circularity. $\eta=1.0$ in all cases.}
\label{fig:goodj2}
\end{figure}

Fig.~\ref{fig:tmerge} shows the merging time-scales as function of the
initial mass ratio, $\mu_{\rmi}$, for three different values of the
orbital circularity, as indicated in each panel. In all cases the
initial orbital energy has $\eta = 1.0$, and we have adopted
\abest{}.  The solid lines correspond to the predictions of model M2
using the above Coulomb logarithm of Equation~(\ref{eq:fitting}), while the
dashed lines indicate the fitting formula of BK08.  Although not
perfect, our model is in fair agreement with the simulation results of
BK08. This is also evident from Fig.~\ref{fig:goodj2}, where we show
the evolution of orbital angular momentum for six different
combinations of $\mu_{\rmi}$ and $\varepsilon$ (as indicated). In each
panel the dashed curve corresponds to the simulation results of BK08,
while the solid line is our M2 model prediction based on the Coulomb
logarithm of Equation~(\ref{eq:fitting}). The corresponding value of $C$ is
indicated in each panel. Overall, our model yields $j(t)$ that are in
satisfactory agreement with the simulation results of BK08.

\subsection{The Distribution of Subhalo Population}
\label{subsec:subpop}
\begin{figure*}
\centerline{\psfig{figure=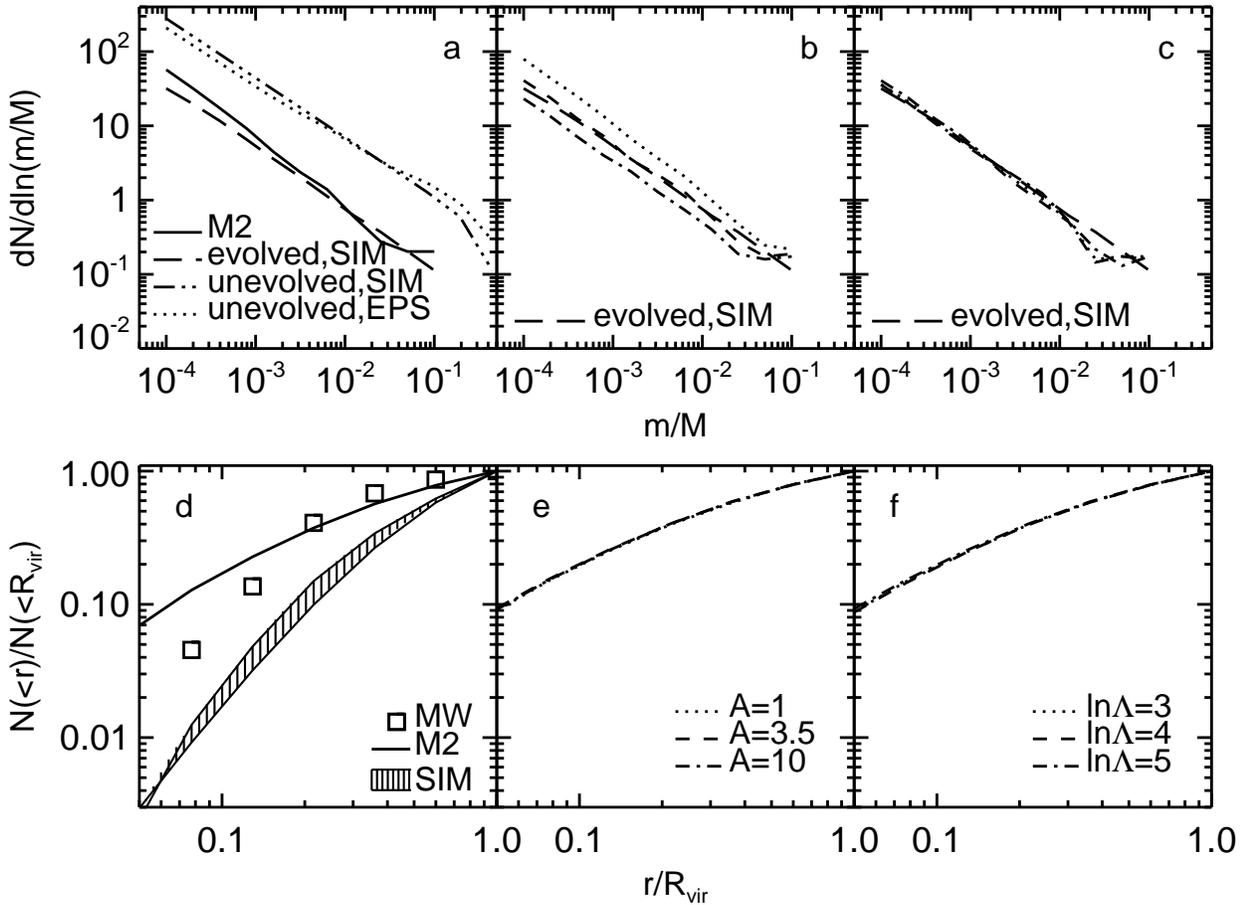,width=0.95\hsize}}
\caption{Subhalo mass function (SHMF,  upper panels) and radial number
  distribution (lower panels) in a  Milky-Way type halo.  Panels a and
  d basically show the predictions from model M2 (solid) with \abest{}
  and \CL{} of Equation~(\ref{eq:fitting}).  In panel a, the unevolved
  and  evolved SHMFs from  simulation are  shown as  dashed-dotted and
  long dashed lines, respectively. In panel d, we also plot the radial 
  number distribution of simulated
  subhaloes  (hatch  area,  upper  limit:  Via  Lactea;  lower  limit:
  Aquarius) and  observational MW  satellites (squares).  In  panels b
  and e,  we compare the  results of simulation  to the model  M1 with
  $\ln \Lambda=-\ln \mu$  and \AAA, while in panels c  and f, the used
  parameters  are  $\ln  \Lambda=3,  4,  5$  and  \abest.   The  model
  predictions  with different  parameters  are plotted  in lines  with
  varying line style as indicated.}
\label{fig:subpop}
\end{figure*}

In Section~\ref{subsec:dynamics}, we have introduced in detail the model 
for the evolution of  subhalo,   including  its   mass, radial  position   
and  merging time-scales. In Section~\ref{subsec:tuning},  we tune the 
model parameters  to fit the
dynamical evolution of subhalo  predicted by simulations. Couple with the
merger trees,  the  model is  ready  to  produce the  subhalo
catalogue in the host halo. As described in Section~\ref{subsec:tree},  
our model employs $100$ realizations of merger trees of the MW type halo, 
and each realization specifies a random assembly history of dark matter 
haloes (Fig.~\ref{fig:MAH}). We follow the dynamical evolution of the 
accreted subhaloes [with masses $m(z_{\rm acc}) \ge 10^8 M_{\odot}$] by 
the main branch of merger tree, and investigate the
distribution of subhalo population at $z=0$, including the subhalo
mass function (SHMF)  and their radial distribution. We also compare the
model prediction with the simulation  results and  the observed  
distribution of  the MW satellites.

Fig.\ref{fig:subpop} show the SHMF  and the radial number distribution
in the upper and lower panels, respectively. For panels in each column,
the same  set of model parameters  is used and indicated  in the lower
panel.  The SHMF from N-body  simulations is well described by a power
law, with  index between $-0.8$  and $-1.0$ (Springel \etal  2001; Gao
\etal 2004b; Kang \etal 2005; Diemand \etal 2004; Giocoli \etal 2008;
2010).  In  panel a,  we show  the simulated one  from Giocoli  et al.
(2008) as  the long-dashed line and  the model prediction  is shown as
the solid line. It can be  seen that our fiducial model (Model M2 with
$A=3.5$ and Coulomb logarithm from Equation~(\ref{eq:fitting})) produces
a fair match to the simulation result.

As the SHMF is for an  evolved population of accreted subhaloes, it is
important to check  if the unevolved SHMF, which  is the mass function
of subhaloes at their accretion  times, is reproduced by the EPS based
merger  tree  employed  in   our  model.   The  unevolved  SHMFs  from
simulation and  the EPS  model are shown  as dashed-dotted  and dotted
lines, respectively. Their good agreement indicates that the model for
the  dynamical evolution  of subhalo  is not  biased by  the formation
history of the host halo.

We  further  check  if  our  model predictions  are  affected  by  the
assumptions for the dynamical processes  of subhalo. Panel b and c show
the predictions from our Model M1, with the dependence on $A$ (panel b)
and Coulomb  logarithm (panel c).  It  is found that  the SHMF depends
strongly on tidal stripping efficiency  $A$, but weakly on Coulomb 
logarithm. This
can be understood  from that, as shown by van  den Bosch \etal (2005),
the subhaloes  population at  present day is  dominated by  the recent
(the last  $\sim 1-2$ Gyr) accretion  history of the host  halo. It is
already  shown  in  Fig.~\ref{fig:modelA}  that subhalo  mass  depends
strongly on $A$  at the first few Gyrs, but with  a weak dependence on
Coulomb logarithm. Thus the results indicate that SHMF can not be used
to constrain the mass evolution of subhalo after a few Gyrs, while the
dynamical evolution $j$  can set strong constraints on  the late stage
evolution of subhalo, as shown in Section~\ref{subsec:tuning}.

The  radial number  distribution of  subhaloes is  shown in  the lower
panels of  Fig.\ref{fig:subpop}.  In panel  d, the hatched  area shows
the spanned  distribution from  simulations (upper limit:  Via Lactea;
lower limit: Aquarius). The observed distribution of the MW satellites
is shown  as the empty  squares (data are  from Mateo 1998;  Kroupa et
al. 2005; Metz et al. 2007; Metz et al.  2009; Martin et al. 2008).  A
clear discrepancy  is that  the distribution of  the MW  satellites is
more concentrated  than the  subhaloes from N-body  simulations, which
has  already been  noted before  (e.g. Taylor  et al.  2005b).   Such a
discrepancy  could  be  due  to  the  incompleteness  of  observations
(Willman et  al.  2004), or  the observed satellites present  a biased
population of subhaloes from simulations (Kravtsov et al.  2004b; Madau
et al.  2008).  We leave more discussion to Section~\ref{sec:discuss}.

The fiducial model  prediction is shown as the solid  line in panel d.
Compared to the simulation result, the model predicts a more centrally
concentrated  distribution of  subhaloes.  A  Similar  discrepancy was
also noted by  Taylor \& Babul (2005b) although  their model prediction
is  slightly lower  than ours.   However, Z05  found that  their model
predicts a well  match to simulation result, and  they argued that the
discrepancy  noted by  Taylor \&  Babul (2005b)  is not  from numerical
effects of  simulation but the  model assumptions for  subhalo merging
and disruption.  There still lacks  detailed studies on  this issue.
Here we firstly explore if  the predicted distribution of subhaloes is
affected by the  model assumption.  The predictions from  our Model M1
are shown in  panel e and panel f, with dependence  on $A$ and Coulomb
logarithm,  respectively.  Surprisingly,  we find  that  the predicted
distribution is  similar to  that obtained from  our Model M2,  and it
also has no dependence on the model parameter $A$ and $\ln \Lambda$.

In principle,  the final spatial  distribution of subhaloes  is mainly
determined by (1) their  initial positions at accretion, (2) dynamical
processes governing  subhalo evolution, and (3) criteria  on where and
when subhalo disappears. The results in panel e and f suggest that (2)
has no  significant effects on  the radial distribution  of subhaloes.
Since the low-mass subhaloes dominate the subhaloes population, varying 
the strength of the dynamical friction and tidal stripping will not 
change the spatial distribution of subhaloes much. 
In addition, Kang  (2008) has shown that the  formation history of the
host halo from the EPS theory  is very similar to that of simulations.
As   the   mass   and   radius   of  halo   are   close   related   by
Equation~(\ref{eq:delta}),  the   initial  positions  of  subhaloes at
accretion (the  virial radius of host  halo) from the  EPS merger tree
should be similar to the simulation results. Thus effect (1) will also
contribute little to the  discrepancy on the final radial distribution
of subhaloes.

It is then reasonable to conclude that the over-predicted subhaloes at
small  radii   is  because   either  simulations  still   lack  enough
resolutions to  resolve subhaloes  in the central  region of  the host
halo,   or   the  model   neglecting   subhalo   disruption  is   not
realistic. With  respect to simulation,  Springel et al.   (2008) have
shown  that  increasing  the  resolution does  resolve  more  low-mass
subhaloes,  but  the number  of  subhaloes  converges  for given  mass
limit.  Thus   it  is   implausible  that  simulation   resolution  is
responsible for this discrepancy.  With respect to the model, defining
a  subhalo  to be  disrupted  (or  unbound)  is very  subjective,  for
example, most authors assume that  subhalo is tidally disrupted if its
mass is less than the initial mass within a radius $f_{dis}r_{s}$, but
with  a wide range  of $f_{dis}$  between $0.1\sim  2.0$. As  shown by
Wetzel \&  White (2010), varying $f_{dis}$  has a huge  impacts on the
final radial distribution of subhaloes.

In fact,  there are  more effects which  can affect the  abundance of
subhaloes  and  their  radial   distribution. 
(i) Host  halo  formed  in
cosmological simulation always contains more than one subhalo, and the
interaction  between  subhaloes  will  accelerate  the  disruption  of
subhaloes and reduce  the number of subhaloes at  the inner host halo
(e.g., Tormen et al. 1998; Gnedin et al. 2004;  Angulo et al. 2009). 
(ii) Ludlow  et  al.   (2009)  have shown  that  small
subhaloes are more likely to be ejected out to larger distances during
the virialization of the host halo, thus producing a less concentrated
distribution.  
(iii) Subhalo-subhalo mergers may be  also effective to reduce the 
abundance of subhaloes (e.g., Kim et al. 2009), especially for the less 
massive  subhaloes (e.g., Angulo et al. 2009).
Unfortunately,  these processes  are  difficult to  be
included in the analytical model.

\begin{figure*}
\centerline{\psfig{figure=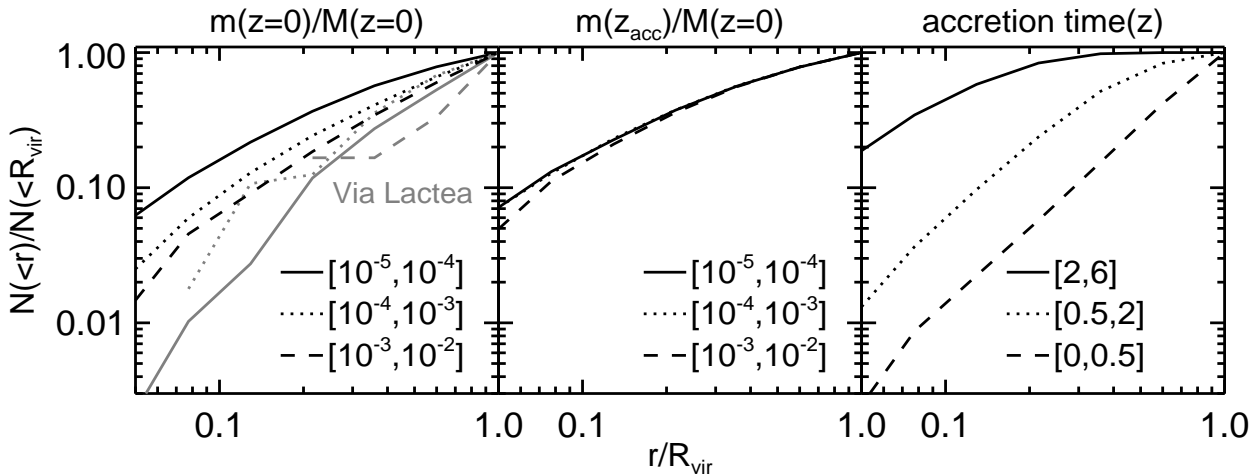,width=0.95\hsize}}
\caption{The  dependence of subhalo radial distribution on   subhalo  
  properties.  The  left,  middle and  right  panels show  the
  dependence on  the present-day mass, mass at  accretion of subhaloes
  (both in unit of present host halo mass, as indicated in each panel)
  and their accretion redshift.
  In  left panel, it also shows the distribution from the data of ``Via
  Lactea'' simulation as in grey lines.}
\label{fig:rfunc}
\end{figure*}

Finally in this  section we consider the dependence  of subhalo radial
distribution on their properties.   Most N-body simulations have shown
that the  radial distribution  of subhaloes has  no dependence  on the
present-mass of  subhaloes (e.g.,  Gao et al.   2004b; Diemand  et al.
2004;  Springel et  al.  2008),  while others
(e.g.,  De Lucia  et  al.  2004)  found  non-negligible dependence  on
subhalo  mass.   In Fig.~\ref{fig:rfunc}  we  show  the fiducial  model
predictions with dependence on the present-day mass (left panel), mass
at accretion (middle panel) and accretion redshift (right panel).  The
subhaloes  mass are  in  unit of  host  halo mass  at  $z=0$, and  are
indicated in  each panel. In  the left panel,  the grey lines  are the
results   obtained   from   the   public  data   of   ``Via   Lactea''
simulation\footnote{http://www.ucolick.org/$\sim$diemand/vl}, where
only the dependence on subhalo present-day mass can be derived.

The  left  panel  shows  that  the  radial  distribution  has  a  weak
dependence on the present-day mass of subhalo. This is consistent with
the  results of  De Lucia  et al.   (2004).  Interestingly,  the ``Via
Lactea'' simulation show an  opposite trend that higher-mass subhaloes
have a more concentrated  distribution within $r<0.2R_{vir}$. As ``Via
Lactea'' simulated only one host halo, there is significant scatter on
the  radial distribution  of subhaloes  due to  the limited  number in
given mass  bins.  When split by their initial mass (i.e., mass at 
accretion), the subhaloes have almost the same radial  distribution, as
shown in the middle panel. 
This result also demonstrates that why the ridial distribution of suhaloes
is independent of the dynamical processes (panel e and f of
Fig.~\ref{fig:subpop}), as the subhaloes catalogue is  
dominated by the less massive suhaloes. 
The right panel  shows that there  is a significant
dependence on the age of subhaloes  that the old population has a more
concentrated distribution at small radius. This age-dependence was
also found by Taylor \& Babul  (2005b), and they pointed out that this
is because old subhaloes are accreted at lower distances when the host
halo was smaller at high redshift.

\section{Halo Occupation Distribution of Subhaloes}
\label{sec:HOD}
\begin{figure}
\centerline{\psfig{figure=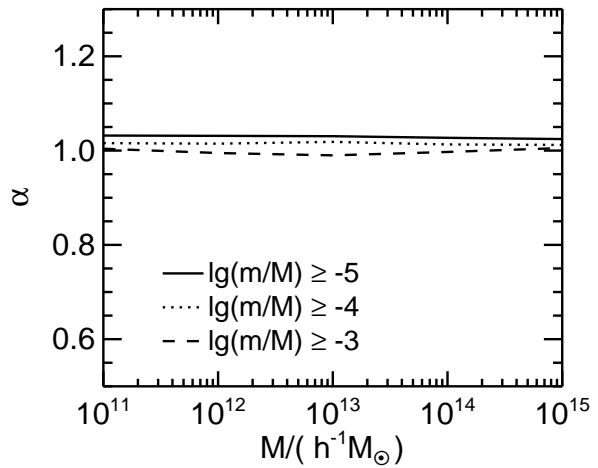,width=0.95\hsize}}
\caption{The normalized second moment of the  HODs of subhaloes as 
  a function of the host halo mass for three
  mass thresholds:  $\lg m/M \geq -5, -4, -3$. }
\label{fig:HOD}
\end{figure}

In order to  further examine the validity of  our model, we investigate
the halo occupation distribution (HOD) of subhaloes. For this purpose,
we  produce subhalo  population  in a  set  of host  haloes with  mass
$M(z=0)=10^{11},      10^{12},      10^{13},     10^{14},      10^{15}
h^{-1}M_{\odot}$. For each host halo, we produce $100$ realizations of 
the   merger  trees,  with   a  resolution   of  $10^7,   10^8,  10^9,
10^{10},10^{11}  M_{\odot}$, respectively.  The fiducial  model  M2 is
used  to model  the  evolution of  each  subhalo.  We count  the
surviving  subhaloes  ($N$)  in  each  realization  and  compute  the
quantity
\begin{equation}
 \label{eq:hod2}
  \alpha \equiv \frac{\left< N(N-1)\right>^{1/2}}{\left< N\right>} 
\end{equation}
for each  host halo.  The quantity $\alpha$  is the  normalized second
moment of subhalo's HOD (e.g., Kravtsov et al. 2004a; van den Bosch et
al. 2005).  For a Poissonian  distribution, $\alpha$ should  be unity,
while  distributions  that are  narrower  (sub-Poissonian) or  broader
(super-Poissonian) have $\alpha < 1$ and $\alpha >1$, respectively. 
In the case of $\alpha \approx 1$ (i.e., small deviation of $\alpha$
from unity), one should keep in mind that it does not necessarily
indicate a Poissonian statistics (e.g., Boylan-Kolchin et al. 2010) and
it needs further examination on the subhalo's HOD. Here we use the
quantity $\alpha$ only for a consistent comparison with the previous
results (e.g., Z05; Zheng et al. 2005).

In Fig.~\ref{fig:HOD}  we plot $\alpha$
as function  of the mass of  the host halo.  We  select subhaloes with
different mass bins of $\lg m/M \geq -5, -4, -3$.  Fig.~\ref{fig:HOD}
shows that $\alpha$ is close to unity independent of the mass threshold
and the host halo mass.
This result is inconsistent with the semi-analytical result
of  van den Bosch  et al.  (2005)  and Z05, who found  that the
distributions of  massive subhaloes in low-mass
host haloes are significantly broader than Poissonian distribution.
They pointed out that the discrepancy may  be from the generic  
problem of  conventional EPS  formalism. However, our model
predictions  agree  with the numerical results
of  Kravtsov et  al. (2004a) and  Zheng et  al. (2005). 
We  argue that  this is due  to the
improvement of our  model in two folds. Firstly,  we employ a modified
EPS formalism by Parkinson et al.   (2008) which is shown to produce a
well  math  to  the  merger   history  of  haloes  found  from  N-body
simulations.   Secondly, we  assume that subhaloes  never get
disrupted  but   are  cannibalized   with  a  time   scale  \dft
(Fig.~\ref{fig:tmerge}).  It gives a better match to the observed
distribution of satellite galaxies. 


\section{Discussion}
\label{sec:discuss}

We have shown in Section~\ref{subsec:subpop} that the model predicts a
more concentrated distribution of subhaloes than that seen from N-body
simulations.  This prediction is  insensitive to the model assumptions
for tidal stripping and dynamical parameters. The same discrepancy was
also obtained by  Taylor \& Babul (2005), and they  ascribed it to the
numerical effects  of simulations.  The  recent high-resolution N-body
simulations (e.g., Springel et al. 2008) have shown that resolution is
not the scapegoat for the low number density of subhaloes in the inner
host halo.   As pointed by Z05 that  it is not the  resolution but the
model  assumption that gives  to the  over-prediction. Z05  obtained a
good  match to  the simulation  result by  implementing  disruption of
subhalo. Wetzel  \& White (2010)  also have shown that  decreasing the
threshold  of tidal disruption  produces more  subhaloes in  the inner
host halo.

We believe  that the good  agreement between our model  prediction and
the observed  distribution of the  MW satellites is a  coincidence. To
get  a better  match  to  the distribution  of  subhaloes from  N-body
simulations,   we should  firstly understand   the  importance of
subhalo-subhalo  interaction  and  subhalo-host  interaction  for  the
disruption of subhaloes, and more studies are needed to classify their
contribution to this discrepancy.  If these interactions are not enough
to dissolve subhaloes  in the inner host halo,  a more realistic model
for  subhalo disruption should  be included  in the  model. Currently,
this  is difficult  to implement  it. In  one aspect,  we need  a more
realistic  and physical  model  for subhalo  disruption, unlike  those
models  to define  disruption when  the  distance of  subhalo to  host
center  is less  than some  given radius  because it  is  arbitrary and
dependent  on  the resolution  of  simulations.   On  the other  hand,
subhalo disruption is easily to  be confused with subhalo merging with
central halo, which is often true for low-resolution simulations.  For
our experiment  in this  paper, we have  to neglect the  disruption and
define subhalo merger using its angular momentum.  This is adopted for
making a fair comparison with the simulations of BK08.

Though we have  neglected subhalo disruption in our  model, the result
indicates that disruption is not  important for real galaxies. This is
seen from panel  d  of  Fig.\ref{fig:subpop}. Indeed,  Hydrodynamical
simulations  with  baryon included  have  confirmed  that subhalo  can
survive the strong  tidal disruption (e.g., Gnedin et  al. 2004; Nagai
\& Kravtsov 2005; Weinberg et al.   2008; Dolag et al.  2009), and the
predicted  spatial distribution  of satellite  galaxies is  similar to
that observed in galaxy clusters (e.g., Gao et al. 2004a). Actually this
is  not the  only  solution to  this  problem. Some  have argued  that
observed satellites  in the MW  are biased tracers of  subhaloes (e.g.,
Kravtsov et al.  2004b; Madau et al. 2008). The  readers are referred to
the review paper by Kravtsov (2010) for more discussions.

\section{Conclusion}
 \label{sec:conc}

In this paper, we study the  evolution of dark matter subhalo using an
analytical  model including  simple descriptions  for a  few important
processes, such as tidal stripping, dynamical friction, tidal heating.
We tune the model parameters to fit the dynamical evolution of subhalo
predicted  by controlled  N-body  simulation.  Then  we combine  these
descriptions with  merger trees from the  EPS-based Monte-Carlo merger
trees  and study  the subhalo  population  in a  Milky-Way type  halo,
including  the subhalo mass  function and  the radial  distribution of
subhaloes.

Following  Boylan-Kolchin  et al.   (2008),  we  define  subhalo to  be
merged with central halo when  its angular momentum reaches zero. We
compare  the predicted  angular momentum  evolution to  the simulation
results of BK08.   We find that the mass loss of  subhalo due to tidal
stripping has great  impact on its angular momentum  evolution. A high
tidal stripping efficiency, $A$, produces a  fast decrease of subhalo  mass and a
longer merger time  scales. We further find that  the mass of subhalo
should  not  decrease  continuously  by tidal  stripping,  and  better
agreement with simulation can be  obtained if the mass of subhalo keep
fixed  after two  passage of  pericenter. We  give a  modified Coulomb
logarithm using  the fitting formula of  BK08 to gain a  well match to
the subhalo merger time-scales.

We compare the subhalo mass  function to that from N-body simulations,
and it is  found that SHMF is mainly determined by  the tidal stripping
efficiency
$A$, but dependent  weakly on the Coulomb logarithm  parameters. It is
also insensitive to subhalo mass loss at its late stage of evolution,
as the  SHMF is dominated by  recently accreted subhaloes.  This is in
good agreement with the results of van den Bosch et al. (2005) and Yang
et al. (2009).

The  radial distribution  of subhaloes  is  found to  be more  central
concentrated  than that  from N-body  simulations.  This  is  a common
prediction  from  analytical  models  if  subhalo  disruption  is  not
important (Taylor \& Babul 2005b;  Z05; Wetzel \& White
2010).   N-body simulations  (e.g., Gao et al. 2004b; 
Springel  et al.   2008)  seems to
indicate that resolution is not the scapegoat for the under-prediction
of subhaloes at small radii.  The interaction between subhalo and host
halo is  efficient to eject small  subhaloes into the  outer region of
host  halo  (Ludlow et  al.   2009)  which  can partly  resolves  this
discrepancy.  To  fully solve  this problem, we  need a  more physical
model for  subhalo disruption  and distinguish between  disruption and
merger, which are not feasible at the moment.

We  conclude  that  the  better agreement  between  predicted  spatial
distribution of subhaloes and  observed satellites in the Milky-Way is
a coincidence. But it also implies that real galaxy may not be tidally
disrupted, as  found from  hydrodynamical simulations (e.g.,  Dolag et
al.   2009)  that condensation  of  baryon  mass  inside subhalo  will
increase the central  density of subhalo, making it  more resistant to
tidal disruption.  Another  argument for the concentrated distribution
of satellite galaxies is that  they are biased population of subhaloes
found from simulations (e.g., Madau  et al. 2008). A final solution to
this issue should turn to the hydrodynamical simulation with high enough
resolution  and  more  realistic  models  for  gas  cooling  and  star
formation in subhaloes.

Finally,  we investigate  the  halo occupation  distribution (HOD)  of
subhaloes with  our improved model.  The second moment of the HODs
is close to unity, which  disagrees with the  results from
semi-analytical  model of  van den  Bosch et  al. (2005)  and  Z05 but
agrees with the  HOD derived from numerical simulation  by Kravtsov et
al. (2004a) and Zheng et al. (2005).

\section*{Acknowledgements}
JLG  acknowledges  the  financial  supports from  Chinese  Academy  of
Sciences  and Max-Planck-Institute  for Astronomy.  We  thanks Stelios
Kazantzidis, Liang Gao and Guoliang Li for helpful discussions.  We also
thanks the referee for useful comments. XK is
supported  by the  {\it One  Hundred Talents}  project of  the Chinese
Academy  of Sciences  and  by the  foundation  for the  author of  CAS
excellent  doctoral dissertation.  JLH  is supported  by the  National
Science Foundation of China under grant No.  10573028, the Key Project
No.  10833005, the  Group Innovation Project No. 10821302,  and by 973
program No. 2007CB815402.

\bibliographystyle{mnras}

\end{document}